\documentclass[aip,preprint,jcp,floatfix]{revtex4-1}
\usepackage{graphicx}
\usepackage{hyperref} 
\usepackage{rotating}
\usepackage{color}
\usepackage{chngcntr}
\usepackage{epstopdf}
\usepackage{braket}
\usepackage{amsmath}
\counterwithin*{paragraph}{subsection}
\definecolor{linkcolour}{rgb}{0,0.2,0.6}
\hypersetup{colorlinks,breaklinks,citecolor=linkcolour,urlcolor=linkcolour,linkcolor=linkcolour}

\begin{document}


\title{Atomic photoionization cross-sections beyond the electric dipole approximation} 
\author{Iulia Emilia Brumboiu}
\email[]{iubr@kth.se}
\affiliation{Department of Theoretical Chemistry and Biology, KTH Royal Institute of Technology, 10691 Stockholm, Sweden}
\affiliation{Department of Chemistry, Korea Advanced Institute of Science and Technology, 34141 Daejeon, Korea}
\author{Olle Eriksson}
\affiliation{Department of Physics and Astronomy, Uppsala University, SE-75120 Uppsala, Sweden}
\author{Patrick Norman}
\affiliation{Department of Theoretical Chemistry and Biology, KTH Royal Institute of Technology, 10691 Stockholm, Sweden}



\date{\today}

\begin{abstract}
A methodology is developed to compute photoionization cross-sections beyond the electric dipole (BED) approximation from response theory, using Gaussian type orbitals and plane waves for the initial and final states, respectively. The methodology is applied to compute photoionization cross-sections of atoms and ions from the first four rows of the Periodic Table. Analysing the error due to the plane wave description of the photoelectron, we find kinetic energy and concomitant photon energy thresholds above which the plane wave approximation becomes applicable. The correction introduced by going beyond the electric dipole approximation increases with photon energy and depends on the spatial extension of the initial state. In general, the corrections are below 10\% for most elements, at a photon energy reaching up to 12 keV. 
\end{abstract}


\maketitle 

\section{Introduction}
Photoionization is the physical process in which atoms, molecules or solids emit electrons upon irradiation. It constitutes the basis for photoelectron spectroscopy (PES), also known as photoemission, an experimental technique routinely used for the characterization of materials.\cite{Hufner2010, Damascelli2003} By recording the kinetic energy of emitted electrons, PES provides a description of the occupied electronic structure of the sample. Besides information on the binding energies of the electronic levels, PE spectra measured using different photon energies can be used to distinguish between states with different atomic compositions.\cite{Green1995, Green2005} 

Commonly, the comparison between experimental PE spectra with calculated electronic structures or total densities of states (DOS) is used to interpret the quality of a particular electronic structure method.\cite{Kummel2008, Kronik2013, Kronik2014, Kronik2016, Egger2014, Marom2012, Kotliar2006, Held2007, Manghi1999} A step beyond such a simple comparison is to include in the calculation the probability of ionization of a particular state. This can be done by multiplying the projected density of states with the corresponding photoionization cross-section calculated for atomic orbitals within the dipole approximation, as performed, for example, in Refs.~[\!\!\citenum{Panda2016, Brena2011, Brumboiu2014, Bidermane2015, Teng2017}]. Another possibility is to compute the dipole transition matrix elements using the Fermi golden rule, as in Refs. [\!\!\citenum{Nguyen2015, Minar2014, Gozem2015, Seabra2004, Seabra2005}]. 

Photoionization cross-sections have been calculated within the dipole approximation for different ranges of photon energies and using different mathematical descriptions for the initial and final states. Most notably, Yeh and Lindau computed the total and partial ionization cross-sections for all atoms in the Periodic Table for photon energies between 10 and 1,500 eV.\cite{Yeh1985, Yeh1993} They used an algorithm introduced by Cooper and Manson,\cite{Cooper1962, Manson1968} where both initial and final states were described on a radial grid and the final state was relaxed in the presence of the positive ion.\cite{Fano1968} Other attempts have also been made using plane waves,\cite{Gozem2015, Seabra2004, Ellison1974, Ellison1975erratum, Deleuze1994} orthogonalized plane waves,\cite{Seabra2004, Ellison1974, Ellison1975erratum, Deleuze1994} Coulomb waves,\cite{Gozem2015, Ritchie1974} or B-splines\cite{Bachau2001, Ruberti2014} to describe the final states. In the solid state community, time-reversed (TR) low energy electron diffraction (LEED) states\cite{Pendry1976, Minar2014, Braun2016} for the final state are used in a Green's function method approach to obtain angle-resolved photoemission spectra (ARPES) of $3d$- and $4f$-materials.\cite{Keqi2018, Veis2018, Gray2011, Minar2014} 

The electric dipole approximation is based on the assumption that the electromagnetic field can be considered uniform across the spatial extent of the initial state wave function. This allows for the truncation of the series expansion of the spatial part of the field at zeroth order, i.e., $e^{i\mathbf{k}\cdot\mathbf{r}}\approx 1$. Here, $\mathbf{k}$ represents the wavevector of the electromagnetic radiation with wavelength $\lambda$, and we have $k=2\pi/\lambda$. Thus, the dipole approximation is not valid when the wavelength $\lambda$ is comparable to or smaller than the spatial extent of the initial-state wave function (or orbital), i.e., when the term $\mathbf{k}\cdot\mathbf{r}$ becomes comparable to or larger than 1. Going beyond the electric dipole approximation becomes especially important when the ionizing photon energy is large and the ionized electronic states are highly delocalized, such as the valence orbitals of conjugated organic molecules or the valence bands of solids. Non-dipole effects have been theoretically described and experimentally observed especially in terms of the angular distributions of the emitted photoelectrons. For overviews from various perspectives, we refer the reader to Refs. [\!\!\citenum{Guillemin2006, Demekhin2014, Lindle1999, Shaw1996, Seabra2005, Grum2003, Derevianko1999, Hemmers1997}] and articles therein.
\newline In the present study, we analyse the correction introduced by using the full field operator in the expression for the atomic photoionization cross-section. For this purpose, we follow the theoretical derivation based on response theory from Ref.~[\!\!\citenum{List2015}] to obtain an equation for the cross-section beyond electric dipole approximation, denoted by $\sigma_{\mathrm{BED}}$ in the present work. By using Gaussian type orbitals (GTOs) and plane waves (PWs) for the initial and final states, respectively, we compute BED cross-sections for the atoms in the first four rows of the Periodic Table, alongside their most common ions, and estimate the magnitude of the BED correction as a function of photon energy for each system. Furthermore, we analyse the error introduced by the PW description for the outgoing photoelectron and find a kinetic energy threshold above which this approximation becomes applicable. 

The article is organized as follows: the full derivation of the expression for $\sigma_{\mathrm{BED}}$ is presented in Section \ref{Theory}, the computational details are described in Section \ref{Details}, while results for selected atoms are presented in Section \ref{Results}. The complete set of data for all atoms and ions included in this study is provided at \url{https://static.sys.kth.se/cbh/teochem/}.

\section{Theory}\label{Theory}
In this section we present a detailed derivation of the equations used to calculate atomic and molecular photoionization cross-sections beyond the electric dipole approximation. First, we define the electronic Hamiltonian which describes the interaction between the atomic or molecular system with a classical electromagnetic wave. Second, we derive a general equation for the absorption cross-section and, finally, we obtain an expression for the photoionization cross-section by approximating the wave function corresponding to the photoelectron by a plane wave (PW) and by describing the atomic orbitals as a linear combination of Gaussian type orbitals (GTOs). 
\subsection{The Interaction Hamiltonian}
The interaction of a system of electrons within an atom or molecule with a time-dependent electromagnetic field, assuming the Born-Oppenheimer approximation, can be described by the non-relativistic Hamiltonian:\cite{Zettili2009, Helgaker2012, Schatz1993}
\begin{equation}
\hat{H}(t)=V(\mathbf{r}_1,\mathbf{r}_2,...,\mathbf{r}_N)+\sum_{j=1}^{N}\left\{\frac{1}{2m_{\mathrm{e}}}\left[\hat{\mathbf{p}}_j+e\mathbf{A}(\mathbf{r}_j,t)\right]^2-e\phi(\mathbf{r}_j,t)+\frac{e}{m_{\mathrm{e}}}\mathbf{B}(\mathbf{r}_j,t)\cdot\hat{\mathbf{s}}_j\right\} ,
\end{equation}
where $N$ is the number of electrons, $m_{\mathrm{e}}$ is the electron mass, $e$ is the elementary charge, $\hat{\mathbf{p}}_j=-i\hbar\boldsymbol{\nabla}_j$ is the momentum operator, $\hat{\mathbf{s}}_j$ is the spin operator, $\mathbf{A}(\mathbf{r}_j,t)$ is the time-dependent vector potential, $\phi(\mathbf{r}_j,t)$ is the time-dependent scalar field and $\mathbf{B}(\mathbf{r}_j,t)$ is the time-dependent magnetic field.
\newline In the Coulomb gauge, for radiation without electrostatic sources, the vector potential is chosen divergence free and the constant scalar potential is imposed to vanish (i.e. $\boldsymbol{\nabla}\cdot\mathbf{A}$=0 and $\phi(\mathbf{r},t)=0$).\cite{Schatz1993} Additionally, the Zeeman term is small compared to the others and may be neglected. By also neglecting the term proportional to $\mathbf{A}^2(\mathbf{r}_j,t)$ as it does not contribute to one-photon processes,\cite{MolecElectrodyn} the equation becomes:
\begin{equation}
\hat{H}(t)=\hat{H}_0+\hat{V}(t) , \label{hamiltonian}
\end{equation}
where 
\begin{align}
\hat{H}_0=V(\mathbf{r}_1,\mathbf{r}_2,...,\mathbf{r}_N)+\sum_{j=1}^{N}\frac{1}{2m_{\mathrm{e}}}\hat{\mathbf{p}}^2_j , \nonumber\\ \hat{V}(t)=\frac{e}{m_{\mathrm{e}}}\sum_{j=1}^{N}\mathbf{A}(\mathbf{r}_j,t)\cdot\hat{\mathbf{p}}_j .\nonumber 
\end{align}
In the Coulomb gauge, for a classical monochromatic electromagnetic wave with angular frequency $\omega$, the vector potential, the electric field, and respectively the magnetic field are:\cite{List2015}
\begin{gather}
\mathbf{A}(\mathbf{r}_j,t)=A_0\boldsymbol{\epsilon}\left[e^{i\mathbf{k}\cdot\mathbf{r}_j-i\omega t}+e^{-i\mathbf{k}\cdot \mathbf{r}_j+i\omega t} \right]=\mathbf{A}_j^\omega e^{-i\omega t}+\mathbf{A}_j^{-\omega} e^{i\omega t}=\sum_{\omega_1=\pm\omega}\mathbf{A}_j^{\omega_1}e^{-i\omega_1t} ,\label{vector_potential}\\
\mathbf{E}(\mathbf{r}_j,t)=-\frac{\partial\mathbf{A}}{\partial t}=\sum_{\omega_1=\pm\omega}i\omega_1\mathbf{A}_j^{\omega_1}e^{-i\omega_1 t} ,\label{electric_field}\\
\mathbf{B}(\mathbf{r}_j,t)=\boldsymbol{\nabla_j}\times\mathbf{A}=i(\mathbf{k}\times\boldsymbol{\epsilon})A_0\left[e^{i\mathbf{k}\cdot \mathbf{r}_j-i\omega t}-e^{-i\mathbf{k}\cdot \mathbf{r}_j+i\omega t} \right] ,\label{magnetic_field}
\end{gather}
where $\boldsymbol{\epsilon}$ is the unit vector of the polarization, $\mathbf{k}$ is the wavevector in the direction of propagation, $\mathbf{A}^{\omega_1}(\mathbf{r}_j)\equiv\mathbf{A}_j^{\omega_1}$ implicitly contains the dependence on $\mathbf{r}_j$, and $A_0=\mathcal{E}_0/2i\omega$ in terms of the real amplitude of the electric field, $\mathcal{E}_0$. We note that the intensity of the incoming wave may be calculated from the time-averaged magnitude of the Poynting vector over one period,\cite{Griffiths1962} i.e. $I(\omega)=\varepsilon_0\,c\,\mathcal{E}_0^2/2$, with $\varepsilon_0$ the vacuum permittivity.
\newline Finally, the interaction potential of Eq.~\eqref{hamiltonian} may be written as:
\begin{equation}
\hat{V}(t)=\sum_{j=1}^{N}\left[\hat{\nu}^{\omega}_j\,e^{-i\omega t}+\hat{\nu}^{-\omega}_j\,e^{i\omega t}\right]=\hat{V}^\omega\,e^{-i\omega t}+\hat{V}^{-\omega}\,e^{i\omega t}=\sum_{\omega_2=\pm\omega}\hat{V}^{\omega_2}e^{-i\omega_2 t} , \label{potential}
\end{equation}
where we have denoted:
\begin{equation}
\hat{\nu}_j^{\omega} =\frac{e}{m_{\mathrm{e}}}\mathbf{A}_j^{\omega}\cdot\hat{\mathbf{p}}_j=-\frac{e\hbar\mathcal{E}_0}{2\omega m_{\mathrm{e}}}e^{i\mathbf{k}\cdot \mathbf{r}_j}\left(\boldsymbol{\epsilon}\cdot\boldsymbol{\nabla}_j\right) .\label{v_omega_plus}
\end{equation}
%
Using Eq.~\eqref{potential}, the Hamiltonian takes the form:
\begin{equation}
\hat{H}(t)=\hat{H}_0+\sum_{\omega_2=\pm\omega}\hat{V}^{\omega_2}e^{-i\omega_2 t} .
\end{equation} 
\subsection{Absorption Cross-Sections from Response Theory}
The problem presented in the previous section may now be addressed using response theory, as discussed in detail in Ref.~[\!\!\citenum{Olsen1985}]. Briefly, response theory consists in describing the time evolution of the particle system $\ket{0(t)}$ using a unitary transformation of the time-independent ground state $\ket{0}$, where the time-dependent unitary operator $\hat{P}(t)$ is expanded in series in terms of the perturbation $\hat{V}(t)$. The evolution of the average value of a generic operator $\hat{O}$ can be calculated using $\ket{0(t)}$. Specific response functions may be obtained by truncating the series at the desired order. A detailed derivation is provided in the Supplementary Material. Given that the process we aim to describe is photoionization due to single photon absorption, we focus on the linear response function only, which for a generic operator $\hat{O}$ is:
\begin{align}
\left<\left<\hat{O};\hat{V}^{\omega_2}\right>\right>=-\frac{1}{\hbar}\sum_{n>0}\left(\frac{\bra{0}\hat{O}\ket{n}\bra{n}\hat{V}^{\omega_2}\ket{0}}{\omega_{n0}-\omega_2-i\gamma_n}+\frac{\bra{0}\hat{V}^{\omega_2}\ket{n}\bra{n}\hat{O}\ket{0}}{\omega_{n0}+\omega_2+i\gamma_n}\right) .\label{rsp_fct}
\end{align}
Here, $\ket{n}$ is an excited state of the time-independent Hamiltonian $\hat{H}_0$ with corresponding eigenvalue $E_n$, $\omega_{n0}=(E_n-E_0)/\hbar$, and $\gamma_n$ is the half-width broadening associated with the inverse lifetime of the excited state. 
\newline With the purpose of deriving an equation for the absorption cross-section, we express the rate of absorption within a volume $V$ in terms of the work done by the electric field  $\mathbf{E}(\mathbf{r},t)$ by generating a current density $\mathbf{j}(\mathbf{r},t)$ in the sample (in our case, limited to bound electrons):\cite{Griffiths1962}
\begin{equation}
\Gamma_{i\rightarrow f}=\frac{\mathrm{d}W}{\mathrm{d}t}=\int_{V}\mathbf{E}(\mathbf{r},t)\cdot\mathbf{j}(\mathbf{r},t)\,\mathrm{d}\mathbf{r} .
\end{equation} 
The time average of the absorption rate over one period may be expanded as:\cite{List2015, Mahr1975} 
\begin{align}
\left<\Gamma_{i\rightarrow f}\right>_T=\left<\frac{\mathrm{d}W}{\mathrm{d}t}\right>_T=\alpha(\omega)I(\omega)+\beta(\omega)I^2(\omega)+\gamma(\omega)I^3(\omega) ,\label{absorption_coeff}
\end{align}
where $\alpha$, $\beta$, $\gamma$ are the absorption coefficients and $\sigma(\omega)=\alpha(\omega)$ is the linear absorption cross-section.
\newline To obtain an expression for $\sigma(\omega)$ we use the non-relativistic current density operator: \cite{Bast2009}
\begin{equation}
\hat{\mathbf{j}}(\mathbf{r})=-\frac{e}{2m_{\mathrm{e}}}\sum_{j=1}^{N}\left[\hat{\mathbf{p}}_j\delta(\mathbf{r}_j-\mathbf{r})+\delta(\mathbf{r}_j-\mathbf{r})\hat{\mathbf{p}}_j\right] ,
\end{equation}
where $\delta(\mathbf{r}_j-\mathbf{r})$ is the Dirac delta function and we have neglected the spin contribution and diamagnetic current.
\newline By replacing the generic operator $\hat{O}$ in Eq.~\eqref{rsp_fct} with $\hat{\mathbf{j}}(\mathbf{r})$, we obtain the first order perturbation-induced current density: 
\begin{align}
\mathbf{j}^{(1)}(\mathbf{r},t)
=-\frac{1}{\hbar}\sum_{n>0}\sum_{\omega_2=\pm\omega}\left(\frac{\bra{0}\hat{\mathbf{j}}(\mathbf{r})\ket{n}\bra{n}\hat{V}^{\omega_2}\ket{0}}{\omega_{n0}-\omega_2-i\gamma_n}+\frac{\bra{n}\hat{\mathbf{j}}(\mathbf{r})\ket{0}\bra{0}\hat{V}^{\omega_2}\ket{n}}{\omega_{n0}+\omega_2+i\gamma_n}\right)e^{-i\omega_2 t} .\label{current_rsp}
\end{align}
The first order transition rate may now be calculated using the first order expression of $\mathbf{j}$ from Eq.~\eqref{current_rsp} and the expression for $\mathbf{E}(\mathbf{r},t)$ from Eq.~\eqref{electric_field}:
\begin{flalign}
\int_{V}\mathbf{E}(\mathbf{r},t)\cdot\mathbf{j}^{(1)}(\mathbf{r},t)\,\mathrm{d}\mathbf{r}&=-\frac{i}{\hbar}\sum_{n>0}\,\sum_{\omega_1,\omega_2=\pm\omega}\omega_1\left(\frac{\bra{0}\int_V\mathbf{A}^{\omega_1}(\mathbf{r})\cdot\hat{\mathbf{j}}(\mathbf{r})\,\mathrm{d}\mathbf{r}\ket{n}\bra{n}\hat{V}^{\omega_2}\ket{0}}{\omega_{n0}-\omega_2-i\gamma_n}\right.\nonumber\\
&\left.+\frac{\bra{0}\hat{V}^{\omega_2}\ket{n}\bra{n}\int_V\mathbf{A}^{\omega_1}(\mathbf{r})\cdot\hat{\mathbf{j}}(\mathbf{r})\,\mathrm{d}\mathbf{r}\ket{0}}{\omega_{n0}+\omega_2+i\gamma_n}\right)e^{-i(\omega_1+\omega_2)t}\nonumber\\
&=\frac{ie}{2m_{\mathrm{e}}\hbar}\sum_{n>0}\,\sum_{\omega_1,\omega_2=\pm\omega}\omega_1\left[\frac{\bra{0}\sum_{j=1}^{N}\left(\hat{\mathbf{p}}_j\cdot\mathbf{A}_j^{\omega_1}+\mathbf{A}_j^{\omega_1}\cdot\hat{\mathbf{p}}_j\right)\ket{n}\bra{n}\hat{V}^{\omega_2}\ket{0}}{\omega_{n0}-\omega_2-i\gamma_n}\right.\nonumber\\
&\left.+\frac{\bra{0}\hat{V}^{\omega_2}\ket{n}\bra{n}\sum_{j=1}^{N}\left(\hat{\mathbf{p}}_j\cdot\mathbf{A}_j^{\omega_1}+\mathbf{A}_j^{\omega_1}\cdot\hat{\mathbf{p}}_j\right)\ket{0}}{\omega_{n0}+\omega_2+i\gamma_n}\right]e^{-i(\omega_1+\omega_2)t} ,\label{work}
\end{flalign}
where, in the first step, the order of integration between sample and electronic degrees of freedom is interchanged and, in the second step, the Dirac delta function has been integrated out. 

Considering that in the Coulomb gauge the vector potential is divergence free, Eq.~\eqref{work} can be written
\begin{flalign}
\int_{V}\mathbf{E}(\mathbf{r},t)\cdot\mathbf{j}^{(1)}(\mathbf{r},t)\,\mathrm{d}\mathbf{r}&=\sum_{\omega_1,\omega_2=\pm\omega}\,\,\sum_{n>0}\frac{i\omega_1}{\hbar}\left(\frac{\bra{0}\hat{V}^{\omega_1}\ket{n}\bra{n}\hat{V}^{\omega_2}\ket{0}}{\omega_{n0}-\omega_2-i\gamma_n}+\frac{\bra{n}\hat{V}^{\omega_1}\ket{0}\bra{0}\hat{V}^{\omega_2}\ket{n}}{\omega_{n0}+\omega_2+i\gamma_n}\right)e^{-i(\omega_1+\omega_2)t}\nonumber\\\label{work_V}
&=\sum_{\omega_1,\omega_2=\pm\omega}-i\omega_1G(\omega_1;\omega_2)e^{-i(\omega_1+\omega_2)t} ,
\end{flalign}
\newline where we identified the linear response function:\cite{List2015}
\begin{align}
\left<\left<\hat{V}^{\omega_1};\hat{V}^{\omega_2}\right>\right>=G(\omega_1;\omega_2)=-\frac{1}{\hbar}\sum_{n>0}\frac{\bra{0}\hat{V}^{\omega_1}\ket{n}\bra{n}\hat{V}^{\omega_2}\ket{0}}{\omega_{n0}-\omega_2-i\gamma_n}+\frac{\bra{n}\hat{V}^{\omega_1}\ket{0}\bra{0}\hat{V}^{\omega_2}\ket{n}}{\omega_{n0}+\omega_2+i\gamma_n} .\label{G}
\end{align}
The absorption rate can now be written in terms of the linear response function:
\begin{equation}
\frac{\mathrm{d}W}{\mathrm{d}t}=\sum_{\omega_1,\omega_2=\pm\omega}-i\omega_1G(\omega_1;\omega_2)e^{-i(\omega_1+\omega_2)t} .
\end{equation}
Finally, the time-averaging over one period of time, $T$, is carried out to arrive at
\begin{flalign}
\left<\frac{\mathrm{d}W}{\mathrm{d}t}\right>_T^{(1)}&=\frac{1}{T}\int_{0}^{T}\sum_{\omega_1,\omega_2=\pm\omega}-i\omega_1G(\omega_1;\omega_2)e^{-i(\omega_1+\omega_2)t}\,\mathrm{d}t\nonumber\\
&=\sum_{\omega_1,\omega_2=\pm\omega}-i\omega_1 G(\omega_1;\omega_2)\;\frac{1}{T}\int_{0}^{T}e^{-i(\omega_1+\omega_2)t}\,\mathrm{d}t\nonumber\\
&=i\omega\left[G(-\omega;\omega)-G(\omega;-\omega)\right] ,\label{work_G}
\end{flalign}
where only the terms with $\omega_1=-\omega_2$ of the sum integrate to non-zero values (equal to $T$).
\newline Comparing Eqs.~\eqref{absorption_coeff} and \eqref{work_G}, the linear absorption cross-section can be expressed as:
\begin{equation}
\sigma(\omega)=\frac{i\omega[G(-\omega;\omega)-G(\omega;-\omega)]}{I(\omega)}=\frac{2i\omega[G(-\omega;\omega)-G(\omega;-\omega)]}{\varepsilon_0c\mathcal{E}_0^2} .
\end{equation}
With $G(-\omega; \omega)$ from Eq.~\eqref{G}, we obtain:
\begin{equation}
G(-\omega;\omega)-G(\omega;-\omega)=-\frac{2i}{\hbar}\sum_{n>0}\gamma_n\left(\frac{\bra{0}\hat{V}^{-\omega}\ket{n}\bra{n}\hat{V}^{\omega}\ket{0}}{(\omega_{n0}-\omega)^2+\gamma_n^2}-\frac{\bra{0}\hat{V}^{\omega}\ket{n}\bra{n}\hat{V}^{-\omega}\ket{0}}{(\omega_{n0}+\omega)^2+\gamma_n^2}\right) .\label{GminusG}
\end{equation}
The first term corresponds to photon absorption, while the second term corresponds to stimulated emission. We keep only the absorption term and calculate the absorption cross-section in the limit $\gamma_n\rightarrow 0$:
\begin{flalign}
\lim_{\gamma_n\rightarrow 0 }\sigma(\omega)&=\frac{4\omega}{\hbar\varepsilon_0c\mathcal{E}_0^2}\lim_{\gamma_n\rightarrow 0}\sum_{n>0}\gamma_n\frac{\bra{0}\hat{V}^{-\omega}\ket{n}\bra{n}\hat{V}^{\omega}\ket{0}}{(\omega_{n0}-\omega)^2+\gamma_n^2}\nonumber\\
&=\frac{4\omega}{\hbar\varepsilon_0c\mathcal{E}_0^2}\sum_{n>0}\bra{0}\hat{V}^{-\omega}\ket{n}\bra{n}\hat{V}^{\omega}\ket{0}\lim_{\gamma_n\rightarrow 0}\frac{\gamma_n}{(\omega_{n0}-\omega)^2+\gamma_n^2}\nonumber\\
&=\frac{4\pi\omega}{\hbar\varepsilon_0c\mathcal{E}_0^2}\sum_{n>0}\left|\bra{0}\hat{V}^{-\omega}\ket{n}\right|^2\delta(\omega_{n0}-\omega) ,
\end{flalign}
where we have used:
 \begin{equation}
 \lim_{\gamma\rightarrow 0}\left(\frac{\gamma}{B^2+\gamma^2}\right)=\pi\delta(B) .
 \end{equation}
Note that the dimension of $\delta(\omega_{n0}-\omega)$ is time. 
\newline By further replacing $\hat{V}^{-\omega}$ using Eqs.~\eqref{potential} and \eqref{v_omega_plus}, we obtain:
\begin{flalign}
\lim_{\gamma_n\rightarrow 0}\sigma(\omega)&=\frac{4\pi\omega}{\hbar\varepsilon_0c\mathcal{E}_0^2}\cdot\frac{e^2\hbar^2\mathcal{E}_0^2}{4\omega^2m_{\mathrm{e}}^2}\sum_{n>0}\left|\bra{0}\sum_{j=1}^{N}e^{-i\mathbf{k}\cdot\mathbf{r}_j}\boldsymbol{\epsilon}\cdot\boldsymbol{\nabla}_j\ket{n}\right|^2\delta(\omega_{n0}-\omega)\nonumber\\
&=\frac{\pi\hbar e^2}{\varepsilon_0\,c\,\omega m_{\mathrm{e}}^2}\sum_{n>0}\left|\bra{0}\sum_{j=1}^{N}e^{-i\mathbf{k}\cdot\mathbf{r}_j}\boldsymbol{\epsilon}\cdot\boldsymbol{\nabla}_j\ket{n}\right|^2\delta(\omega_{n0}-\omega) .\label{abs_cross_section}
\end{flalign}
\subsection{Approximations for the Initial and Final States}
Eq.~\eqref{abs_cross_section} is a general expression applicable to single photon absorption where the initial and final states are described by the many-body wave functions $\ket{0}$ and $\ket{n}$, respectively. Considering that the electronic structure of the atoms we are interested in may be approximated by a single Slater determinant, the matrix element of the one-electron operator $e^{-i\mathbf{k}\cdot\mathbf{r}_j}\boldsymbol{\epsilon}\cdot\boldsymbol{\nabla}_j$ can be re-written in terms of spin-orbitals:\cite{Szabo2012}
\begin{equation}
\lim_{\gamma_n\rightarrow 0}\sigma(\omega)=\frac{\pi\hbar e^2}{\varepsilon_0\,c\,\omega m_\mathrm{e}^2}\sum_{f}\left|\bra{\chi_i}e^{-i\mathbf{k}\cdot\mathbf{r}}\boldsymbol{\epsilon}\cdot\boldsymbol{\nabla}\ket{\chi_f}\right|^2\delta(\omega_{fi}-\omega) ,\label{phi_cross_section}
\end{equation}
where $\chi_i$ is an occupied atomic orbital and $\chi_f$ is the wave function corresponding to the photoelectron. We note that the same expression is also valid in the case of molecular orbitals. 
\newline Considering that BED corrections become more important at high photon energies, where the wavelength of the ionizing electromagnetic radiation becomes comparable to the spatial extent of (valence) atomic orbitals, it is reasonable to assume that the photoelectron, with a high kinetic energy, can be described as a periodic plane wave normalized over an arbitrary volume $\Omega$:
\begin{equation}
\chi_f=\frac{1}{\sqrt{\Omega}}e^{i\mathbf{k}_{\mathrm{e}}\cdot\mathbf{r}} ,\label{PW}
\end{equation} 
with density of states:
\begin{equation}
\rho_f(\hbar\omega_f)=\frac{\Omega}{2\pi^2}\frac{(2m_\mathrm{e})^{\frac{3}{2}}}{\hbar^3}\sqrt{\frac{\hbar^2k_\mathrm{e}^2}{2m_\mathrm{e}}}=\frac{\Omega}{2\pi^2}\frac{(2m_\mathrm{e})^{\frac{3}{2}}}{\hbar^3}\sqrt{\hbar\omega_f} . \label{rho}
\end{equation}
Using PWs and the corresponding free electron density of states, in the limit of a large volume $\Omega$, the sum over final states in Eq.~\eqref{phi_cross_section} becomes an integral over a continuum:
\begin{flalign}
\lim_{\gamma_n\rightarrow 0}\sigma(\omega)&=\frac{\pi\hbar e^2}{\varepsilon_0\,c\,\omega m_\mathrm{e}^2}\int_{-\infty}^{\infty}\left|\bra{\chi_i}e^{-i\mathbf{k}\cdot\mathbf{r}}\boldsymbol{\epsilon}\cdot\boldsymbol{\nabla}\ket{\frac{1}{\sqrt{\Omega}}e^{i\mathbf{k}_\mathrm{e}\cdot\mathbf{r}}}\right|^2\rho_f(\hbar\omega_f)\delta(\omega_{f}-\omega_i-\omega)\,\hbar\mathrm{d}\omega_f=\nonumber\\
&=\frac{\pi\hbar e^2}{\varepsilon_0\,c\,\omega m_\mathrm{e}^2}\frac{\hbar}{\Omega}\left|\bra{\chi_i}e^{-i\mathbf{k}\cdot\mathbf{r}}\boldsymbol{\epsilon}\cdot\boldsymbol{\nabla}\ket{e^{i\mathbf{k}_{\mathrm{e}}\cdot\mathbf{r}}}\right|^2\rho_f(\hbar(\omega_i+\omega)) ,
\end{flalign}
where the kinetic energy of the photoelectron ($E_f=\hbar\omega_f$) satisfies energy conservation $\omega_f=\omega_i+\omega$, and the photoelectron wavevector $\mathbf{k}_{\mathrm{e}}$ is correspondingly chosen. The advantage of using a PW for the final state is the fact that PWs are eigenfunctions of the momentum operator and the expression for the photoionization cross-section may be, therefore, further simplified: 
\begin{flalign}
\lim_{\gamma_n\rightarrow 0}\sigma(\omega)&=\frac{\pi\hbar^2 e^2}{\varepsilon_0\,c\,\omega m_\mathrm{e}^2}\frac{1}{\Omega}\left|\bra{\chi_i}e^{-i\mathbf{k}\cdot\mathbf{r}}\boldsymbol{\epsilon}\cdot\boldsymbol{\nabla}\ket{e^{i\mathbf{k}_\mathrm{e}\cdot\mathbf{r}}}\right|^2\frac{\Omega}{2\pi^2}\frac{(2m_\mathrm{e})^\frac{3}{2}}{\hbar^3}\sqrt{\frac{\hbar^2k_\mathrm{e}^2}{2m_\mathrm{e}}}\nonumber\\
&=\frac{e^2}{\pi\varepsilon_0\,c\,m_\mathrm{e}}\frac{k_\mathrm{e}}{\omega}\left|\bra{\chi_i}e^{-i\mathbf{k}\cdot\mathbf{r}}\boldsymbol{\epsilon}\cdot\boldsymbol{\nabla}\ket{e^{i\mathbf{k}_\mathrm{e}\cdot\mathbf{r}}}\right|^2\nonumber\\
&=\frac{e^2}{\pi\varepsilon_0\,c\,m_\mathrm{e}}\frac{k_\mathrm{e}}{\omega}\left|\boldsymbol{\epsilon}\cdot \mathbf{k}_\mathrm{e}\braket{\chi_i|e^{i\mathbf{k}_\mathrm{e}\cdot\mathbf{r}-i\mathbf{k}\cdot\mathbf{r}}}\right|^2 .
\end{flalign}
Finally, by expressing the initial state wave function $\chi_i$ as a linear combination of Gaussian type orbitals (GTOs), the photoionization cross-section becomes:
\begin{equation}
\lim_{\gamma_n\rightarrow 0}\sigma(\omega)=\frac{e^2}{\pi\varepsilon_0\,c\,m_\mathrm{e}}\frac{k_\mathrm{e}}{\omega}\left|\boldsymbol{\epsilon}\cdot \mathbf{k}_\mathrm{e}\sum d_{l}\braket{g_l|e^{i\mathbf{k}_{\mathrm{e}}\cdot\mathbf{r}-i\mathbf{k}\cdot\mathbf{r}}}\right|^2 ,\label{phi_cs_gaussians}
\end{equation}
where the summation is over all primitives used to describe the initial state and $g_l$ is a GTO with $l=l_x+l_y+l_z$:
\begin{align}
g_{l_x, l_y, l_z}=N_l(x-R_x)^{l_x}(y-R_y)^{l_y}(z-R_z)^{l_z}e^{-\alpha\left[(x-R_x)^2+(y-R_y)^2+(z-R_z)^2\right]} ,
\end{align}  
with $R_x$, $R_y$ and $R_z$ the coordinates of the atom where the GTO is centred.
\newline Eq.~\eqref{phi_cs_gaussians} can be straightforwardly generalized to molecular orbitals, by replacing $\chi_i$ with a linear combination of atomic orbitals.
\newline Finally, the matrix element to be calculated is:
\begin{equation}
\braket{g_{lx,ly,lz}|e^{i\mathbf{k}_{\mathrm{e}}\cdot \mathbf{r}-i\mathbf{k}\cdot\mathbf{r}}}=\int\limits_{-\infty}^{\infty}\int\limits_{-\infty}^{\infty}\int\limits_{-\infty}^{\infty}g_{lx,ly,lz}(x,y,z)e^{i\mathbf{K\cdot r}}\,\mathrm{d}x\,\mathrm{d}y\,\mathrm{d}z , \label{Integral}
\end{equation}
where $\mathbf{r}=x\,\mathbf{\hat{e}_x}+y\,\mathbf{\hat{e}_y}+z\,\mathbf{\hat{e}_z}$, and $\mathbf{K}=K_x\,\mathbf{\hat{e}_x}+K_y\,\mathbf{\hat{e}_y}+K_z\,\mathbf{\hat{e}_z}$ determined from the input parameters $\mathbf{k}_{\mathrm{e}}$ and $\mathbf{k}$, $\mathbf{K}=\mathbf{k}_{\mathrm{e}}-\mathbf{k}$.
\newline The advantage of working with PWs and GTOs is the fact that the integrals in Eq.~\eqref{Integral} can be computed analytically.\citep{Straub2009} We provide analytic formulas up to $l=5$ in the Supplementary Material. Furthermore, the dipole approximation is obtained simply by setting the photon wavevector $\mathbf{k}=\mathbf{0}$ and hence the dipole and BED photoionization cross-sections are treated with the same formalism and no further variables are introduced in the comparison. 

\section{Computational Details}\label{Details}
\begin{figure}
 \includegraphics[scale=1.0]{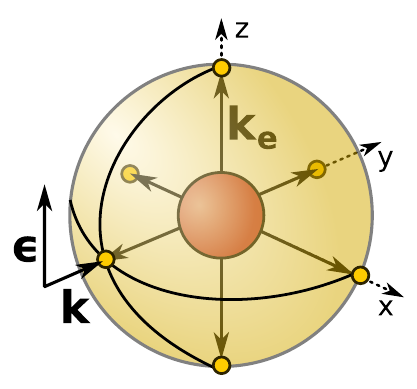}
 \caption{Schematic representation of the computational set-up. The atom is placed at the origin of the coordinate system, the photon wavevector is directed along the y-axis, the polarization direction is along the z-axis and the direction of the photoelectron wavevector is generated using the Lebedev quadrature. The points generated by the quadrature at the 6$^{\mathrm{th}}$ order are marked with yellow dots.}. 
 \label{setup}%
 \end{figure}

The electronic structure of the atoms from the first four rows of the Periodic Table (H-Kr) was calculated with the Gaussian 16\cite{g16} quantum chemistry software, at the B3LYP level of theory\cite{Becke1993} and using the universal Gaussian basis set (UGBS) by de Castro and co-workers.\cite{Silver1978, Silver1978a, Mohallem1986, Mohallem1987, daCosta1987, daSilva1989, Jorge1997, Jorge1997a, deCastro1998} The electronic structure calculations were performed using Cartesian Gaussian functions because the expression for the matrix elements involved in the photoionization cross-section, Eq.~\eqref{phi_cs_gaussians}, has been derived for this GTO variant. For the N atom, we have additionally computed cross-sections using six other basis sets, namely STO-3G,\cite{Hehre1969, Collins1976} 4-31G,\cite{Ditchfield1971, Hehre1972, Hariharan1974, Gordon1980} 6-31G,\cite{Ditchfield1971, Hehre1972, Hariharan1973, Hariharan1974, Gordon1980, Francl1982, Binning1990, Blaudeau1997, Rassolov1998, Rassolov2001} cc-pVTZ,\cite{Kendall1992} and cc-pVQZ.\cite{Woon1993} The basis set results are shown in the Supplementary Material, Fig. S1.  
\newline In addition to the neutral atoms, we have also computed the most common ions corresponding to each atom. The atoms and ions included are listed in Table \ref{Configs}, alongside the electronic configuration relaxed with B3LYP. The relaxed electronic structures were used to compute the dipole and beyond dipole photoionization cross-sections for a range of photon energies between 20 eV to 12 keV, with an energy step of 10 eV. 
\setlength{\tabcolsep}{12pt}
\begin{table}[ht]
\caption{Electronic configuration of the atoms and ions included in this study.}
\begin{tabular}{ccccc}
Z & Element & Configuration & Element & Configuration \\
\hline
1  & H  & $1\mathrm{s}^1$ & & \\
2  & He & $1\mathrm{s}^2$ & &\\
3  & Li & $1\mathrm{s}^22\mathrm{s}^1$ & Li$^{1+}$ & $1\mathrm{s}^2$\\
4  & Be & $1\mathrm{s}^22\mathrm{s}^2$ & Be$^{2+}$ &$1\mathrm{s}^2$\\
5  & B  & $1\mathrm{s}^22\mathrm{s}^22\mathrm{p}^1$ & B$^{3+}$ & $1\mathrm{s}^2$\\
6  & C  & $1\mathrm{s}^22\mathrm{s}^22\mathrm{p}^2$ & &  \\
7  & N  & $1\mathrm{s}^22\mathrm{s}^22\mathrm{p}^3$ & & \\
8  & O  & $1\mathrm{s}^22\mathrm{s}^22\mathrm{p}^4$ & O$^{2-}$ &$1\mathrm{s}^22\mathrm{s}^22\mathrm{p}^6$\\
9  & F  & $1\mathrm{s}^22\mathrm{s}^22\mathrm{p}^5$ & F$^{1-}$ & $1\mathrm{s}^22\mathrm{s}^22\mathrm{p}^6$\\
10 & Ne & $1\mathrm{s}^22\mathrm{s}^22\mathrm{p}^6$ & &\\
11 & Na & [Ne] $3\mathrm{s}^1$ & Na$^{1+}$ & [Ne]\\
12 & Mg & [Ne] $3\mathrm{s}^2$ & Mg$^{2+}$ & [Ne]\\
13 & Al & [Ne] $3\mathrm{s}^23\mathrm{p}^1$ & Al$^{3+}$ & [Ne]\\
14 & Si & [Ne] $3\mathrm{s}^23\mathrm{p}^2$ &  &\\
15 & P  & [Ne] $3\mathrm{s}^23\mathrm{p}^3$ &  &\\
16 & S  & [Ne] $3\mathrm{s}^23\mathrm{p}^4$ & S$^{2-}$ & [Ne] $3\mathrm{s}^23\mathrm{p}^6$ \\
17 & Cl & [Ne] $3\mathrm{s}^23\mathrm{p}^5$ & Cl$^{1-}$ & [Ne] $3\mathrm{s}^23\mathrm{p}^6$\\
18 & Ar & [Ne] $3\mathrm{s}^23\mathrm{p}^6$ & &\\
\end{tabular}
\label{Configs}
\end{table} 
\setcounter{table}{0}
\begin{table}
\caption{(Continued) Electronic configuration of the atoms and ions included in this study.}
\begin{tabular}{ccccc}
Z & Element & Configuration & Element & Configuration\\
\hline
19  & K  & [Ar] $4\mathrm{s}^1$ & K$^{1+}$ & [Ar]\\
20  & Ca & [Ar] $4\mathrm{s}^2$ & Ca$^{2+}$ & [Ar]\\
21  & Sc & [Ar] $4\mathrm{s}^23\mathrm{d}^1$ & Sc$^{3+}$ & [Ar] \\
22  & Ti & [Ar] $4\mathrm{s}^23\mathrm{d}^2$ & Ti$^{4+}$ & [Ar] \\
23  & V  & [Ar] $4\mathrm{s}^23\mathrm{d}^3$ & V$^{5+}$ & [Ar] \\
24  & Cr & [Ar] $4\mathrm{s}^13\mathrm{d}^5$ & Cr$^{6+}$ & [Ar] \\
25  & Mn & [Ar] $4\mathrm{s}^23\mathrm{d}^5$ & Mn$^{2+}$ & [Ar] $3\mathrm{d}^5$ \\
26  & Fe & [Ar] $4\mathrm{s}^23\mathrm{d}^6$ & Fe$^{2+}$ & [Ar] $3\mathrm{d}^6$ \\
27  & Co & [Ar] $4\mathrm{s}^23\mathrm{d}^7$ & Co$^{2+}$ & [Ar] $3\mathrm{d}^7$ \\
28  & Ni & [Ar] $4\mathrm{s}^23\mathrm{d}^8$ & Ni$^{2+}$ & [Ar] $3\mathrm{d}^8$ \\
29  & Cu & [Ar] $4\mathrm{s}^13\mathrm{d}^{10}$ & Cu$^{2+}$ & [Ar] $3\mathrm{d}^{9}$ \\
30  & Zn & [Ar] $4\mathrm{s}^23\mathrm{d}^{10}$ & Zn$^{2+}$ & [Ar] $3\mathrm{d}^{10}$ \\
31  & Ga & [Ar] $4\mathrm{s}^23\mathrm{d}^{10}4\mathrm{p}^1$ & Ga$^{3+}$ & [Ar] $3\mathrm{d}^{10}$\\
32  & Ge & [Ar] $4\mathrm{s}^23\mathrm{d}^{10}4\mathrm{p}^2$ & &  \\
33  & As & [Ar] $4\mathrm{s}^23\mathrm{d}^{10}4\mathrm{p}^3$ & &  \\
34  & Se & [Ar] $4\mathrm{s}^23\mathrm{d}^{10}4\mathrm{p}^4$ & Se$^{2-}$ & [Ar] $4\mathrm{s}^23\mathrm{d}^{10}4\mathrm{p}^6$ \\
35  & Br & [Ar] $4\mathrm{s}^23\mathrm{d}^{10}4\mathrm{p}^5$ & Br$^{1-}$ & [Ar] $4\mathrm{s}^23\mathrm{d}^{10}4\mathrm{p}^6$\\
36  & Kr & [Ar] $4\mathrm{s}^23\mathrm{d}^{10}4\mathrm{p}^6$ & &\\
\end{tabular}
\end{table}
A photoionization cross-section was calculated for each atomic orbital in the relaxed electronic configuration using Eq.~\eqref{phi_cs_gaussians} and the computational set-up shown in Figure \ref{setup}. The wavevector of the photon, $\mathbf{k}$, was set in the y-direction, while the polarization vector was chosen in the z-direction. The norm of $\mathbf{k}$ is calculated from the photon energy $\hbar\omega$. The norm of the photoelectron wavevector $\mathbf{k}_{\mathrm{e}}$ is calculated from the kinetic energy $\mathrm{KE}$, in turn, computed from energy conservation:
\begin{equation}
\mathrm{KE}=\hbar\omega+\varepsilon_{\mathrm{AO}}\, \mbox{; } \, k_{\mathrm{e}}=\frac{\sqrt{2m_{\mathrm{e}}\,\mathrm{KE}}}{\hbar} ,
\end{equation}
where $\varepsilon_{\mathrm{AO}}$ represents the computed eigenvalue of a particular atomic orbital.
\newline The BED photoionization cross-section is finally computed by averaging over all possible photoelectron directions generated on a sphere using the Lebedev quadrature.\cite{Lebedev1976} We tested the Lebedev quadrature at different orders. Due to the spherical symmetry of the atomic system, the lowest order of the quadrature (i.e. 6) sufficed in the case of the dipole photoionization cross-section. For BED, the order 50 was instead required to fully converge the cross-sections. We have therefore computed both BED and dipole cross-sections using the order 50 for the Lebedev quadrature.
\newline In the case of open-shell atoms, we additionally averaged over the possible initial state configurations of the partially filled shell. For example, in the case of the C atom, the 2p cross-section is computed as follows:
\begin{equation}
\sigma_\mathrm{2p}=\frac{\sigma_{\mathrm{p^\alpha_xp^\alpha_y}}+\sigma{_\mathrm{p^\alpha_yp^\alpha_z}}+\sigma_{\mathrm{p^\alpha_xp^\alpha_z}}}{3} ,
\end{equation}
where the index $\mathrm{p^\alpha_x}$ means that the spin-up $2\mathrm{p_x}$ orbital is occupied.
\newline The total atomic photoionization cross-section was calculated by summing over the cross-sections of the occupied atomic shells. Dipole photoionization cross-sections were calculated by the same algorithm with the only difference that the wavevector $\mathbf{k}$ was set to zero.
\section{Results and Discussion}\label{Results}
In the following, (Section A) we analyse the performance of the PW approximation for the final state by comparing our calculated photoionization cross-sections to calculated results from the literature and experimentally measured data. (Section B) We then compare the dipole and BED cross-sections to each other and calculate the magnitude of the relative correction introduced by going beyond the electric dipole approximation as a function of photon energy. 

\subsection{Performance of the PW Approximation}
\begin{figure}[ht]
\includegraphics[scale=1.0]{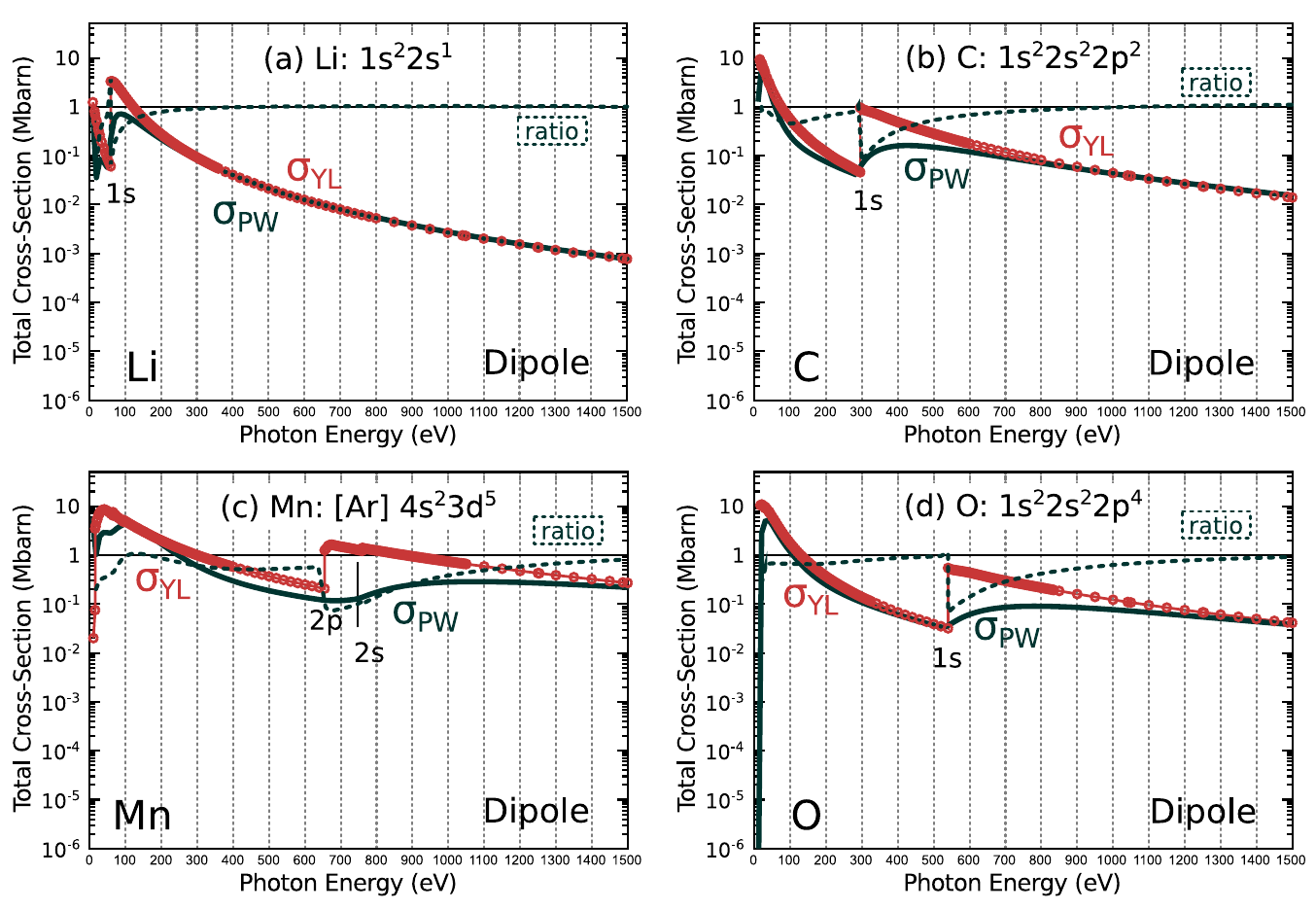}
\caption{Comparison between total dipole photoionization cross-sections calculated using a PW final state ($\sigma_{PW}$, in dark green) and total dipole photoionization cross-sections calculated by Yeh and Lindau\cite{elettra, Yeh1985, Yeh1993} ($\sigma_{YL}$, in red) for (a) Lithium, (b) Carbon, (c) Manganese, and (d) Oxygen. The ratio between $\sigma_{PW}$ and $\sigma_{YL}$ is also depicted (dotted line). The labels 1s, 2s, and 2p mark the ionization thresholds for these orbitals. The PW cross-sections have been multiplied by a factor of 1/3 to average over the directions of the incoming photon, as performed in Refs. [\!\!\citenum{Yeh1985, Yeh1993}]. Note that the vertical axis is logarithmic.}
\label{PW_vs_YL}
\end{figure}

Figure \ref{PW_vs_YL} shows the total dipole photoionization cross-section calculated using a PW final state ($\sigma_{\mathrm{PW}}$) in comparison to the cross-section calculated by Yeh and Lindau\cite{elettra, Yeh1985, Yeh1993} using a final state relaxed in the presence of the positive ion ($\sigma_{\mathrm{YL}}$). The comparison is shown as a function of photon energy, from 20 eV up to 1500 eV, for Li (Fig. \ref{PW_vs_YL}a), C (Fig. \ref{PW_vs_YL}b), Mn (Fig. \ref{PW_vs_YL}c), and O (Fig. \ref{PW_vs_YL}d). The figure also depicts the ratio $\sigma_{\mathrm{PW}}/\sigma_{\mathrm{YL}}$ as a function of photon energy. 
\newline What becomes clear from the figure is that $\sigma_{\mathrm{PW}}$ converges towards $\sigma_{\mathrm{YL}}$ as the photon energy is increased, i.e. as the kinetic energy of the photoelectron increases. In the vicinity of the ionization threshold, the PW approximation performs poorly, but it improves as the photon energy is increased. The photon energy above which the PW approximation becomes applicable depends on the particular element, becoming larger as the atomic number increases. For example, going approximatively 100 eV above the 1s threshold is required in the case of the Li atom, but more than 700 eV above the 2s threshold is instead required for the Mn atom. The poor performance in the vicinity of the ionization threshold is expected, since the outgoing electron, in this case, has low kinetic energy and is not free electron like.

\begin{figure}[ht]
\includegraphics[scale=1.0]{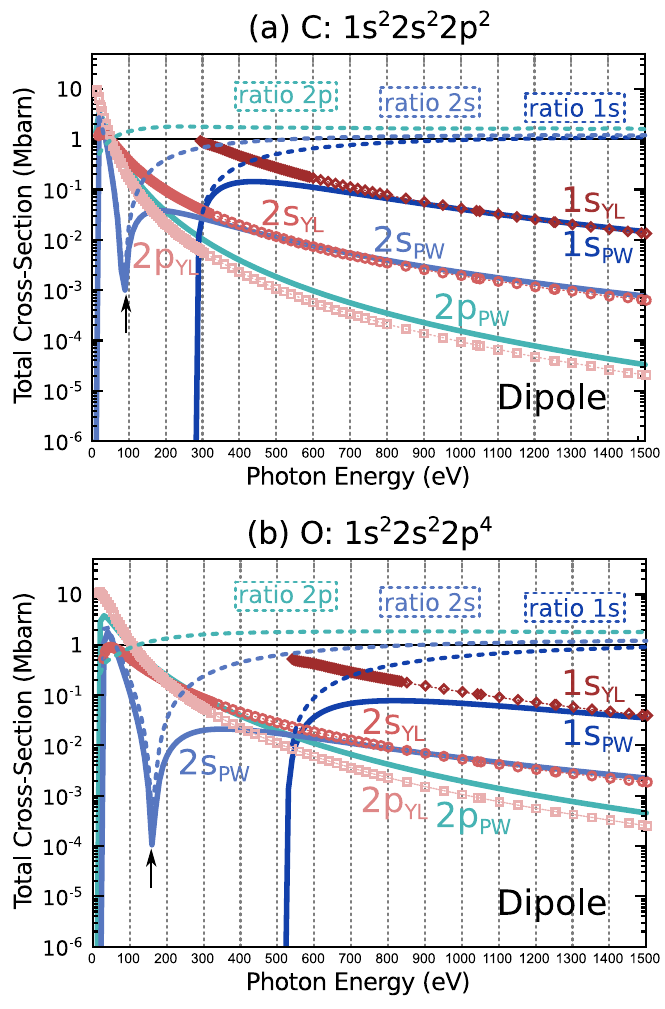}
\caption{Comparison between partial dipole photoionization cross-sections calculated using a PW final state (blue solid lines) and partial dipole photoionization cross-sections calculated by Yeh and Lindau\cite{elettra, Yeh1985, Yeh1993} (red symbols) for (a) Carbon, and (b) Oxygen. The ratio between $\sigma_{PW}$ and $\sigma_{YL}$ is also depicted (blue dotted lines). The PW cross-sections have been multiplied by a factor of 1/3 to average over the directions of the incoming photon as performed in Refs. [\!\!\citenum{Yeh1985,Yeh1993}]. Note that the vertical axis is logarithmic. The inset shows the product between the radial parts of a 2s atomic orbital with a PW as a function of the distance to the nucleus (r).}
\label{PW_vs_YL_partial}
\end{figure}

\begin{figure}[ht]
\includegraphics[scale=1.0]{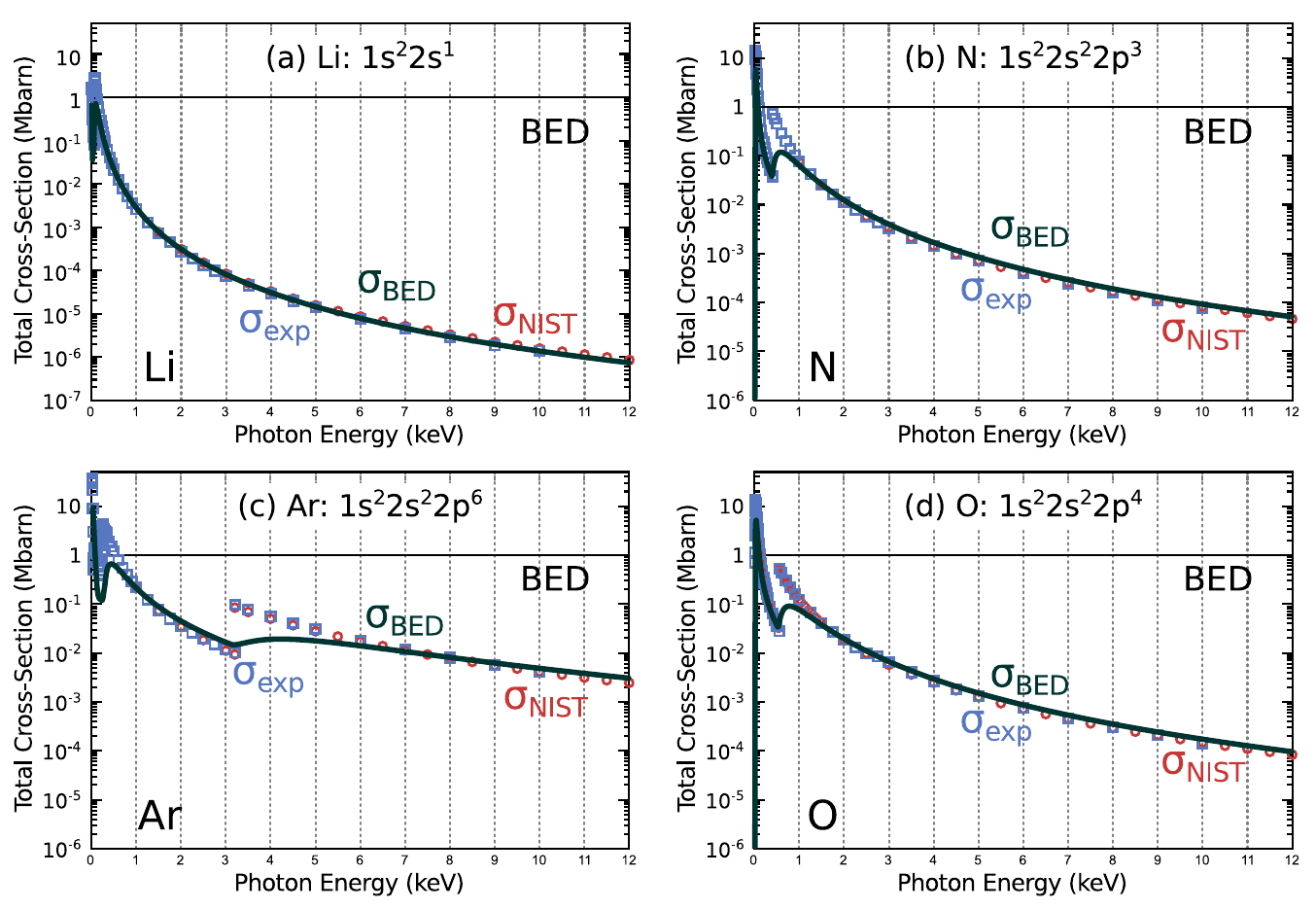}
\caption{Comparison of the total BED photoionization cross-sections calculated using a PW final state ($\sigma_{\mathrm{BED}}$, black line) with measured photoionization cross-sections compiled by Berkowitz\cite{Berkowitz2002} (stored on the NIFS database,\cite{NIFS} $\sigma_{\mathrm{exp}}$, blue squares), and with total dipole cross-sections calculated using the Hartree-Fock-Slater method ($\sigma_{\mathrm{NIST}}$, red circles) from the NIST database.\cite{NIST} The comparison is shown for (a) Li, (b) N, (c) Ar, and (d) O. $\sigma_{\mathrm{BED}}$ has been multiplied by a factor of 1/3 to average over the directions of the incoming photon, as performed in Refs. [\!\!\citenum{Yeh1985, Yeh1993}]. Note that the vertical axis is logarithmic.}
\label{PW_vs_exp}
\end{figure}

In addition, we have also analysed selected partial dipole photoionization cross-sections and how they compare to the results obtained by Yeh and Lindau.\cite{elettra, Yeh1985, Yeh1993} These comparisons are shown in Figure \ref{PW_vs_YL_partial} for C (Fig. \ref{PW_vs_YL_partial}a), and O (Fig. \ref{PW_vs_YL_partial}b) in the photon energy window between 20 eV and 1500 eV. As it can be seen from Fig. \ref{PW_vs_YL_partial}, the PW approximation becomes better further away from the ionization threshold, but different atomic orbitals behave differently. The agreement to the Yeh and Lindau results\cite{elettra,Yeh1985,Yeh1993} is better in the case of s orbitals than in the case of p orbitals. 
\newline It is interesting to note that the cross-sections calculated for the 2s orbitals of both the C and O atoms  present a pronounced dip, at $\hbar\omega\approx 90$ eV and $\hbar\omega\approx 160$ eV, respectively  (marked by arrows in Figure \ref{PW_vs_YL_partial}). The reason for this dip is that the transition matrix element is zero at that particular photon energy. Because the 2s atomic orbital has one node, it is expressed as a linear combination of GTOs with both positive and negative coefficients. Since the integrals involving s-type GTOs and a PW are always real and positive (see the analytical formulas for the integrals in the Supplementary Material), there are both negative and positive contributions to the transition matrix element. For a particular value of the photoelectron wavevector, the positive and negative contributions cancel out, giving rise to the sharp dip in the photoionization cross-section. All other orbitals which have nodes present similar dips. General equations for different types of atomic orbitals described as linear combinations of GTOs may be worked out. We include one example for a very simple 2s orbital in the Supplementary Material. We should note that such discontinuities would not arise for Slater type orbitals (STOs), as exemplified in the Supplementary Material for a 2s STO.  
\newline To further elucidate if the PW approximation for the final state is applicable, we have compared the beyond electric dipole cross-section ($\sigma_{\mathrm{BED}}$) to experimental data and calculations performed over a much larger photon energy range. Figure \ref{PW_vs_exp} shows this comparison for photon energies in the range of 20 eV to 12 keV for Li (Fig. \ref{PW_vs_exp}a), N (Fig. \ref{PW_vs_exp}b), Ar (Fig. \ref{PW_vs_exp}c), and O (Fig. \ref{PW_vs_exp}d). The experimental data was compiled for nine atoms from selected reference sources by Berkowitz.\cite{Berkowitz2002} The measurements are typically performed in gas phase and at room temperature, as described in Refs.~[\!\!\citenum{Schmidt2005, Samson1994, Berkowitz1979}]. The calculated data was obtained within the dipole approximation using the Hartree-Fock-Slater method (the same algorithm used by Yeh and Lindau) and is stored on the National Institute of Standards and Technologies (NIST) database.\cite{NIST} 
\newline The figure shows that far from the ionization threshold the cross-sections calculated using a PW final state (beyond the electric dipole approximation) compare very well with the experimental data. Additionally, the values of $\sigma_{\mathrm{BED}}$ are very close to the calculated dipole cross-sections $\sigma_\mathrm{NIST}$. As it will be discussed in the following section, the reason for this similarity is that the relative correction introduced by BED to the total photoionization cross-section does not amount to more than $\sim$5\%.

\begin{figure}
\includegraphics[scale=1.0]{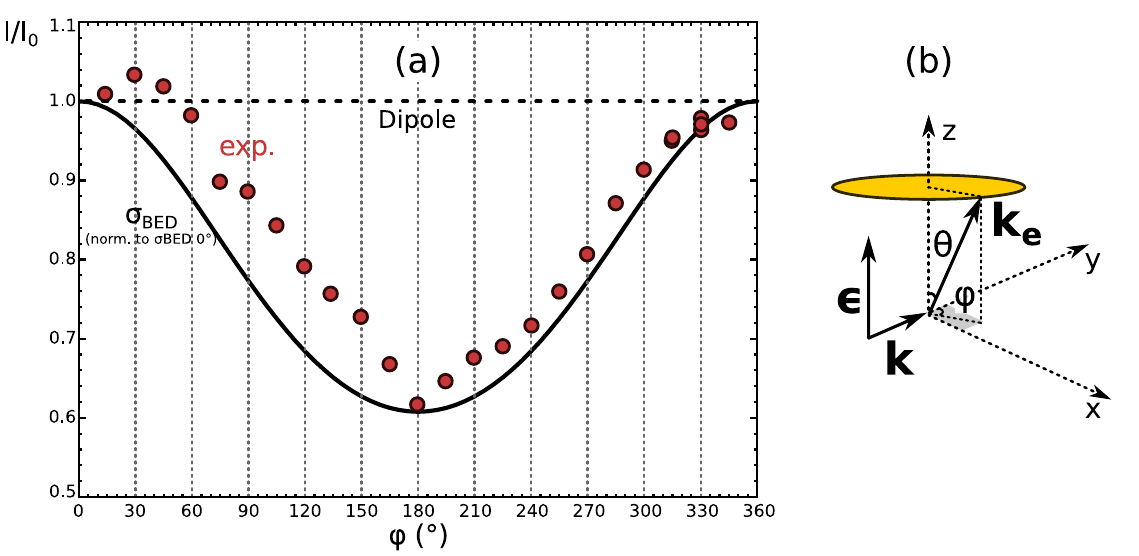}
\caption{Comparison between the angular distributions of photoelectrons calculated using the beyond electric dipole cross-sections (black solid line) and the experimental photoelectron yield (red circles) as a function of the azimuthal angle $\varphi$. The angular distribution is determined for the 1s orbital of Ar and a photon energy $\hbar\omega=5206$ eV. The experimental data is from Ref. [\!\!\citenum{Krassig1995}]. The values on the y-axis have been normalized to the cross-section calculated, respectively measured, at $\varphi=0^{\circ}$. The inset depicts the experimental set-up, where the analyser is placed at a fixed polar angle $\theta=54.7^{\circ}$ and rotated on the yellow circle to vary $\varphi$.}
\label{AngDistrib}
\end{figure}

We have also computed the photoelectron angular distribution at a photon energy $\hbar\omega=5206$ eV, for the Ar 1s orbital, shown in Figure \ref{AngDistrib}. The angular distribution of photoelectrons is typically measured using linearly polarized light by placing an electron analyser at a fixed polar angle $\theta$, as depicted in Figure \ref{AngDistrib}b. The number of electrons emitted at different azimuthal angles is then recorded by rotating the electron analyser around the z-axis and performing measurements at different values of $\varphi$. If the photoionization process would take place according to the dipole approximation, the number of emitted photoelectrons would not depend on $\varphi$ and the graph in Fig. \ref{AngDistrib}a would be a straight horizontal line. The deviation of the curve from a straight line is therefore a measure of nondipolar asymmetry. The differential cross-section for photoionization ($\mathrm{d}\sigma/\mathrm{d}\Omega$) using linearly polarized light, taking into account quadrupole terms, may be expressed as:\cite{Krassig1995,Cooper1993}
\begin{equation}
\frac{\mathrm{d}\sigma}{\mathrm{d}\Omega}=\frac{\sigma}{4\pi}\left[1+\beta P_2(\cos{\theta})+(\delta+\gamma\cos^2{\theta})\sin{\theta}\cos{\varphi}\right] ,
\end{equation}
where $\Omega$ is the solid angle, $\beta$ is the anisotropy parameter, $\delta$ and $\gamma$ are the asymmetry parameters which quantify the nondipolar effects, and $P_2(\cos{\theta})$ is the second Legendre polynomial. 
\newline At the magic angles $\theta=54.7^{\circ}$ and $180^{\circ}-\theta$, $P_2(\cos{\theta})$ is zero and the differential cross-section then depends only on the azimuthal angle $\varphi$:\cite{Jung1996}
\begin{equation}
\frac{\mathrm{d}\sigma}{\mathrm{d}\Omega}=\frac{\sigma}{4\pi}\left[1+\sqrt{\frac{2}{27}}(\gamma+3\delta)\cos{\varphi}\right] .
\end{equation}  

We have calculated $\sigma_{BED}$ for the 1s atomic orbital of Ar at different $\varphi$ angles by rotating the $\mathbf{k}_\mathrm{e}$ vector around the z-axis, as depicted in Figure \ref{AngDistrib}b. The photon polarization was kept in the z direction and the wavevector of the photon was kept in the y-direction. We normalized the value of the calculated cross-section to the value calculated for $\varphi=0^\circ$, as performed for the experimental data.\cite{Krassig1995} The experimental and calculated curves match rather well, and our calculated angle-dependent $\sigma_{\mathrm{BED}}$ cross-section is able to reproduce the magnitude of nondipolar effects observed experimentally.

To summarize this section, by comparing our atomic photoionization cross-sections calculated using a PW final state to previously calculated results from the literature and experimental data, we can conclude that the PW approximation for the final state is applicable at high kinetic energies, i.e. far from the ionization threshold. However, the kinetic energy above which the approximation is applicable depends on the atom and type of orbitals, since the comparison becomes worse for larger atoms and for orbitals with larger $l$ quantum number. Using the PW final state, the cross-sections computed for orbitals with one or more nodes have discontinuities at photon energies (generally close to the ionization threshold) where the positive and negative contributions to the transition matrix element cancel out. These are inherent to the GTOs used for the initial state. Finally, our calculated $\sigma_{\mathrm{BED}}$ is able to reproduce the nondipolar angular distribution of photoelectrons at high photon energies. 

\begin{figure}[ht]
\includegraphics[scale=1.0]{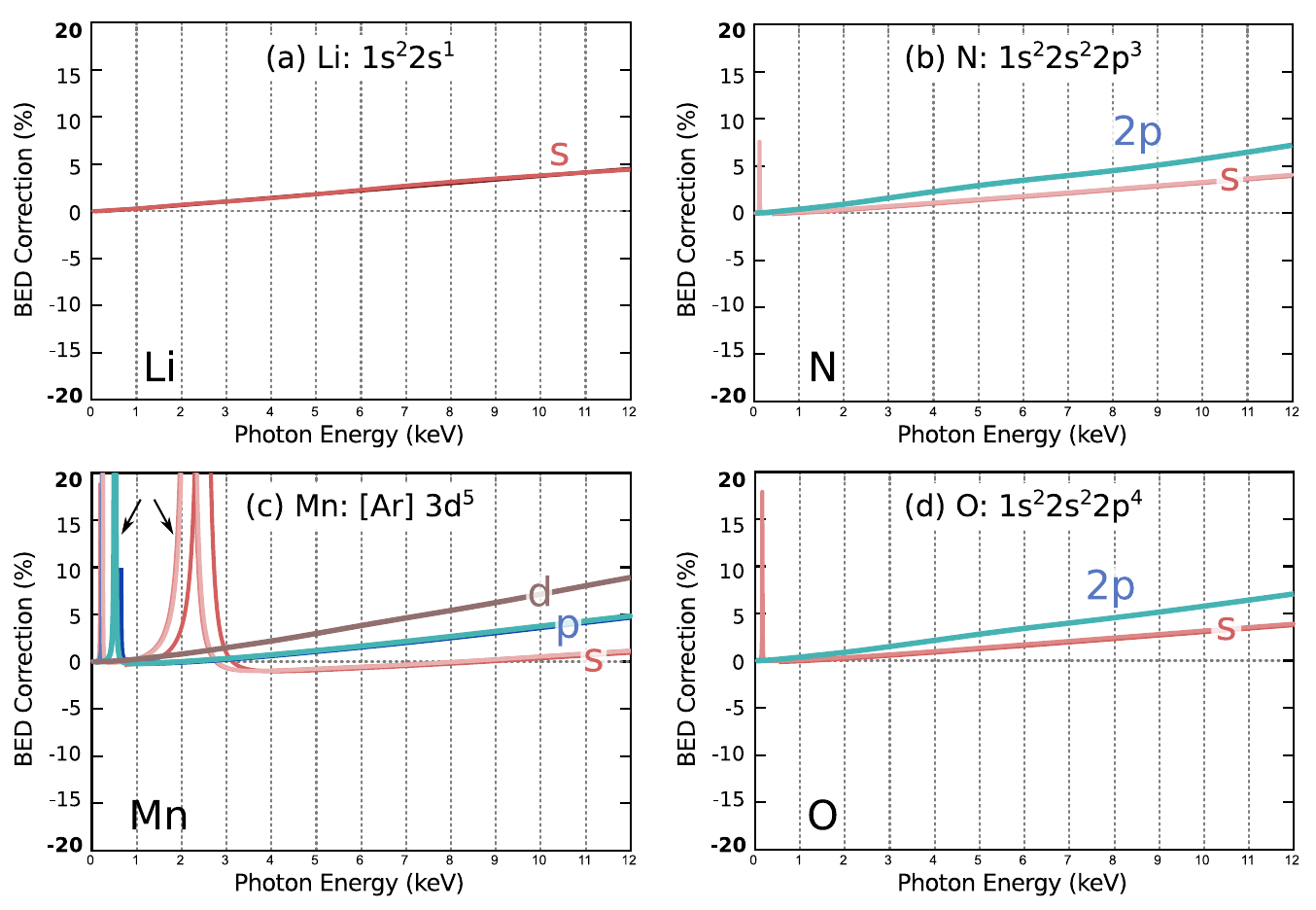}
\caption{The magnitude of the correction introduced by going beyond the dipole approximation as a function of photon energy for (a) Li, (b) N, (c) Mn, and (d) O. The corrections to partial cross-sections are represented in pink (s orbitals), blue (p orbitals), and brown (d orbitals).}
\label{Corrections}
\end{figure}

\subsection{Magnitude of the BED Correction}
Having analysed the advantages and limitations of a PW final state, we now move to examine the magnitude of the correction introduced by going beyond the electric dipole approximation. Figure \ref{Corrections} shows the correction  obtained for total and partial photoionization cross-sections of selected atoms as a function of photon energy, between 20 eV and 12 keV. The correction is represented as:
\begin{equation}
\Delta\sigma=100\frac{\sigma_{\mathrm{BED}}-\sigma_{\mathrm{dipole}}}{\sigma_{\mathrm{BED}}}\,\, [\%]
\end{equation}
The partial cross-sections are grouped together based on the orbital type (s, p, d) because the values computed from different shells (1s, 2s, etc.), but the same $l$ quantum number, are very similar to each other. 
\newline As expected, the BED correction increases as the photon energy is increased. In the case of the s partial cross-sections, the values reach only up to 5\% for a photon energy $\hbar\omega=12$ keV. This is not unexpected given that only at this photon energy the wavelength of the photon (1.95 a.u.) begins to become comparable to the size of the orbitals ($\sim$ 2 a.u., for a 2s orbitals). The total cross-sections (not shown) follow closely the behaviour of the s-type orbitals. This is also expected, since the last shell to ionize will be the one which will dominate, as illustrated in Figure \ref{PW_vs_YL_partial}, where the 1s cross-section is 2-3 orders of magnitude larger than the cross-sections of the 2s and 2p orbitals. For all the elements of the first four rows of the Periodic Table, at 12 keV the last shell to ionize is either a 1s or a 2s shell, resulting in the high similarity between the correction for s partial photoionization cross-section and the total one. 
\newline In the case of p and d atomic orbitals, the correction introduced by BED are slightly larger, between 5-10\% for the three atoms shown in Fig. \ref{Corrections}. One last aspect to note is the appearance of sharp dips, generally at low photon energies, notably visible in the case of the Mn atom (marked with arrows in Fig. \ref{Corrections}c). These are related to the discontinuities previously discussed, resulting from the cancellation between the positive and negative contributions to the transition matrix element which involves an orbital with at least one node. 

\begin{figure}[ht]
\includegraphics[scale=1.0]{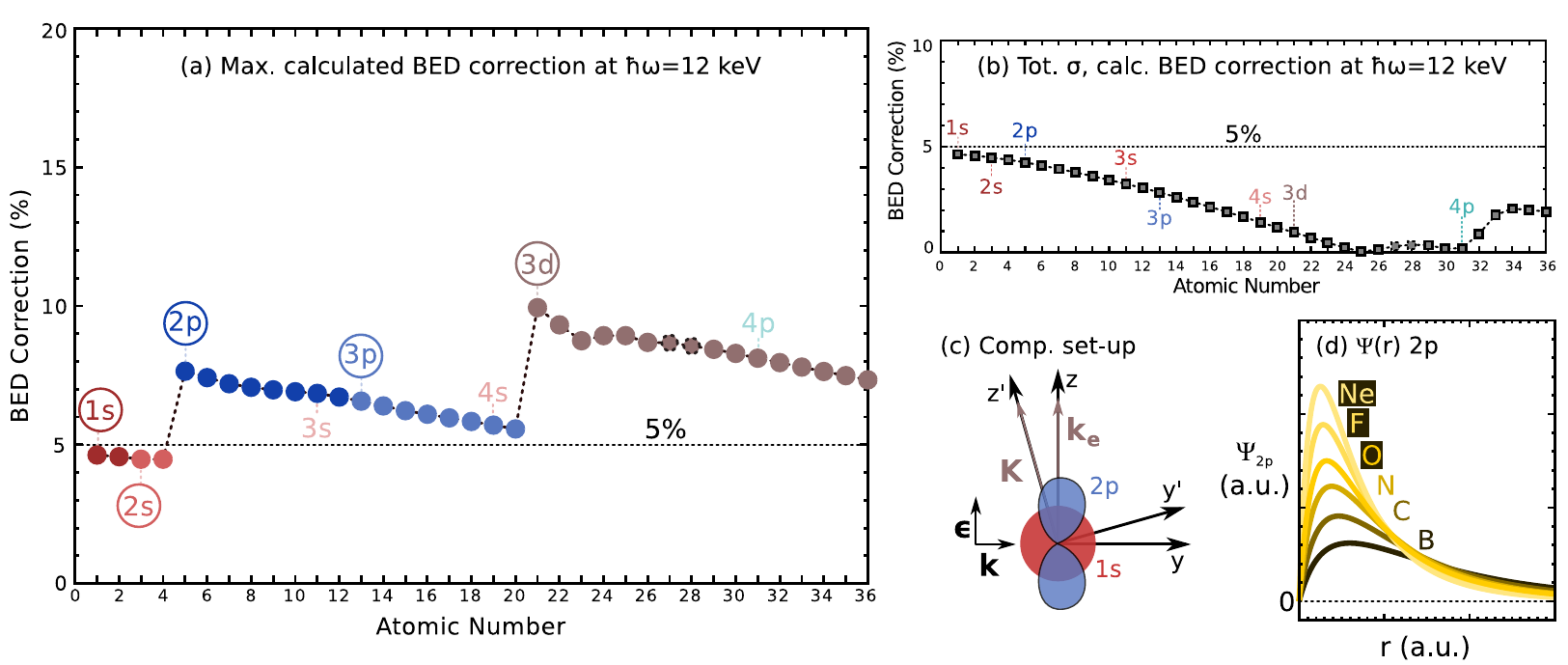}
\caption{The magnitude of the correction introduced by going beyond the dipole approximation as a function of the atomic number, at a photon energy $\hbar\omega=12 \mathrm{keV}$, for (a) the maximum calculated corrections to partial cross-sections, and (b) the corrections to total cross-sections. For the points marked with dotted lines the data shown was computed for the ions instead of the neutral atoms.\footnote{For the Ni and Co atoms, the B3LYP electronic configuration relaxed to a situation where the 4s orbital remained unoccupied irrespective of the initial guess. We have, therefore, used the calculated data for the ions instead.} The corrections to partial cross-sections in (a) are coloured according to the orbital shell they were calculated for. (c) The computational set-up used to calculate the partial cross-sections, and (d) the radial part of the 2p$_\mathrm{z}$ orbital as calculated using the B3LYP functional and the UGBS basis set for atoms B to Ne.}
\label{Corrections_all}
\end{figure}

Finally, we collected the corrections introduced by going beyond the electric dipole approximation for all atoms at the largest photon energy included in the study, i.e. $\hbar\omega=12$ keV. The results are represented as a function of atomic number in Figure \ref{Corrections_all}. The corrections to the total cross-section are depicted in gray (Fig \ref{Corrections_all}b), while the l-shells with the largest correction to the cross-sections are depicted in color (Fig. \ref{Corrections_all}a). The dots are red if the maximum correction was obtained for an s orbital, blue for a p orbital, and brown for a d orbital. The atomic number at which a new shell starts to fill is marked by a dotted line. Note that the maximum correction may be obtained for a different orbital than the last to be filled. For example, the largest correction obtained for the Na atom corresponds to a photoionization of the 2p orbitals, rather than the 3s which is the last to be filled. 
\newline The first feature to notice in Figure \ref{Corrections_all}a is that there are three ``edges": the first at Z=1, the second at Z=5, and the third at Z=21. Interestingly, these edges arise when making a transition from $l$ to $l+1$, i.e. going from s to p, and, respectively, from p to d, where the corrections to the p and d cross-sections are larger.
\newline The larger sensitivity of the p and d orbitals to the BED correction may be understood by visualizing the effect of the BED correction in the computational set-up, as represented in Figure \ref{Corrections_all}c. The dipole approximation is obtained by setting the photon wavevector to zero, meaning the transition matrix elements that have to be calculated involve the initial state (s,p or d orbital), and a PW of wavevector $\mathbf{k}_\mathrm{e}$ directed along the z-axis. Going beyond the dipole approximation ($\mathbf{k}\neq 0$) translates into having to calculate transition matrix elements between the initial state and a new PW now directed along the z'-axis, with a new $\mathbf{K}=\mathbf{k}_\mathrm{e}-\mathbf{k}$. This new PW has both a different magnitude and a different direction, compared to $\mathbf{k}_{\mathrm{e}}$. In the case of the s orbital, which is spherically symmetric, only the difference in magnitude affects the transition matrix element, while in the case of the p and d orbitals, both magnitude and direction affect the transition matrix element, making this type of orbitals more sensitive to the dipole approximation. 
\newline Another feature of the graph in Figure \ref{Corrections_all}a is that each edge is followed by a linear decrease of the magnitude of the BED correction. This decrease may be explained by the decrease in the spatial extent of the particular type of orbitals when going through the series. This is illustrated for the 2p elements in Figure \ref{Corrections_all}d.  
\section{Conclusions}
In summary, we have derived an equation to compute photoionization cross-sections beyond the electric dipole approximation from response theory, using gaussian type orbitals for the initial state and a plane wave with appropriate kinetic energy for the final state. We compared our calculated cross-sections to experimental data and to calculated dipole photoionization cross-sections from the literature. This comparison showed that the PW approximation for the final state is applicable far from the ionization threshold. However, how far is sufficient depends on the atom and the type of atomic orbital, the comparison becoming worse for the larger atoms and for orbitals with larger $l$ quantum number. 
\newline Given that the dipole and beyond dipole photoionization cross-sections can be computed analytically and on the same footing, we could estimate the relative correction introduced by going beyond the electric dipole approximation. As expected, this correction increases with photon energy, but it is, in general, rather small, $\sim$5-10\% at the largest photon energy we included (12 keV). The fact that we obtain such a small correction is not surprising, considering the small spatial extent of the atomic orbitals. Even though the BED corrections for the atoms are quite small, we expect larger corrections for more delocalized states as, for example, molecular orbitals of conjugated molecules, or valence states of solids.
\newline Finally, the description for the atomic photoionization cross-sections may be improved by replacing the PW final state with a linear combination of PWs relaxed in the presence of the positive ion. This is, however, rather complicated because the relaxation should be performed for a small kinetic energy window centred around the kinetic energy obtained from energy conservation. Considering all states up to this kinetic energy is simply not feasible.

\section*{Supplementary Material}
The supplementary material contains \textbf{a)} the complete derivation of the linear response function corresponding to a time-dependent perturbation $\hat{V}$ and a generic operator $\hat{O}$; \textbf{b)} analytical formulas for the integrals in Eq.~\eqref{Integral} \textbf{c)} photoionization cross-sections calculated using different basis sets for N; \textbf{d)} the analytical form of a transition matrix element between a simplified 2s orbital (described as a linear combination of two GTOs) and a plane wave;

\begin{acknowledgments}
\noindent Financial support from the Knut and Alice Wallenberg Foundation (Grant No.{\textbackslash} KAW-2013.0020) and the Swedish Research Council (Grant No.{\textbackslash} 621-2014-4646) is acknowledged. The computations were performed on resources provided by the Swedish National Infrastructure for Computing (SNIC) at NSC and HPC2N. I.E.B. is very grateful to Faris Gel'mukhanov, Nanna Holmgaard List, Barbara Brena, and Igor Di Marco for all the insightful discussions. O.E. also acknowledges support from eSSENCE.
\end{acknowledgments}

%

\end{document}